\documentclass[largeformat]{interact}

\usepackage{epstopdf}
\usepackage[caption=false]{subfig}

\usepackage[numbers,sort&compress,merge]{natbib}
\bibpunct[, ]{[}{]}{,}{n}{,}{,}

\theoremstyle{plain}

\theoremstyle{definition}

\theoremstyle{remark}

\begin{document}


\title{Structure and dynamics of HD$^+$ in the vicinity of the H$^+$ + D and D$^+$ + H dissociation thresholds: Feshbach resonances and the role of {\it g/u}-symmetry breaking}

\author{
\name{Maximilian Beyer\textsuperscript{a,b} and Fr\'ed\'eric~Merkt\textsuperscript{a}\thanks{CONTACT Fr\'ed\'eric Merkt. Email: merkt@phys.chem.ethz.ch}}
\affil{\textsuperscript{a} Laboratory of Physical Chemistry, ETH Zurich, Vladimir-Prelog-Weg 2, 8093 Zurich, Switzerland}
\affil{\textsuperscript{b} Department of Physics and Astronomy, Vrije Universiteit, De Boelelaan 1081, 1081 HV Amsterdam, The Netherlands}
}

\maketitle

\begin{abstract}
We report on the measurement and analysis of photoionisation, mass-analysed-threshold-ionisation and pulsed-field-ionisation zero-kinetic-energy photoelectron spectra of HD in the vicinity of the dissociative-ionisation thresholds H$^+$ + D(1s) + e$^-$ and H(1s) + D$^+$ + e$^-$. The spectra were recorded via the $v=11$-13 vibrational levels of the $\bar{\rm H}$ and $\bar{\rm B}$ outer wells of the H$\bar{\rm H}$~$^1\Sigma_g^+$ and B$^{\prime\prime}\bar{\rm B}$~$^1\Sigma_u^+$ states and provide information on the weakly bound levels of the X$^+$ $^2\Sigma_g^+$ ground state of HD$^+$ with vibrational quantum numbers $v^+=16$-21 and rotational angular-momentum quantum numbers $N^+=0$-10. The spectra also reveal for the first time shape and Feshbach resonances of HD$^+$ located energetically above the H$^+$ + D(1s) dissociation limit. The structure and dissociation dynamics of the $\bar{\rm H}$ and $\bar{\rm B}$ states of HD and of the X$^+$ $^2\Sigma_g^+$ and A$^+$ $^2\Sigma_u^+$ states of HD$^+$ near the dissociation thresholds are strongly influenced by the {\it g/u}-symmetry breaking caused by the nuclear-mass asymmetry.
\end{abstract}

\begin{keywords}
PFI-ZEKE photoelectron spectroscopy, shape and Feshbach resonances, $g/u$-symmetry breaking
\end{keywords}

\section{Introduction}\label{intro}

This article presents an experimental study  of HD$^+$ in the vicinity of the H$^+$ + D(1s) and H(1s) + D$^+$ dissociation thresholds by pulsed-field-ionisation zero-kinetic-energy (PFI-ZEKE) photoelectron spectroscopy. We used a resonant multiphoton excitation sequence via the $v=11$-$13$ levels of the H$\bar{\rm H}$~$^1\Sigma_g^+$ and B$^{\prime\prime}\bar{\rm B}$~$^1\Sigma_u^+$ states. 
These levels are located in the outer ($\bar{\rm H}$ and $\bar{\rm B}$) potential wells, which have ion-pair character, and are thus ideally suited to efficiently access the high vibrational levels of HD$^+$ located just below, and the dissociation continua located above, these thresholds. They are almost degenerate and, in HD, the {\it gerade/ungerade (g/u)} symmetry breaking from the nuclear-mass asymmetry mixes these two states \cite{reinhold99a} and makes them accessible by two-photon excitation from the ground state.

The emphasis of the article is placed on (i) the structure and dynamics of HD in the H$\bar{\rm H}$~$^1\Sigma_g^+$ and B$^{\prime\prime}\bar{\rm B}$~$^1\Sigma_u^+$ states and of HD$^+$ near the dissociation threshold, and (ii) the effects of the {\it g/u}-symmetry breaking in the electronic states of HD and HD$^+$. The earlier theoretical work of L. Wolniewicz and his coworkers on these topics \cite{reinhold99a,wolniewicz80a,wolniewicz91a} were extremely useful in guiding our analysis, and so were numerous experimental and theoretical studies of isotopic effects near the dissociation thresholds of HD (see, e.g., Refs.~\cite{thorson71a,dabrowski76a,durup78a,delange00a,delange02a,grozdanov09a,wang18a,wang20a} and references therein).

\begin{figure}
	\centering
	\includegraphics[width=0.5\textwidth]{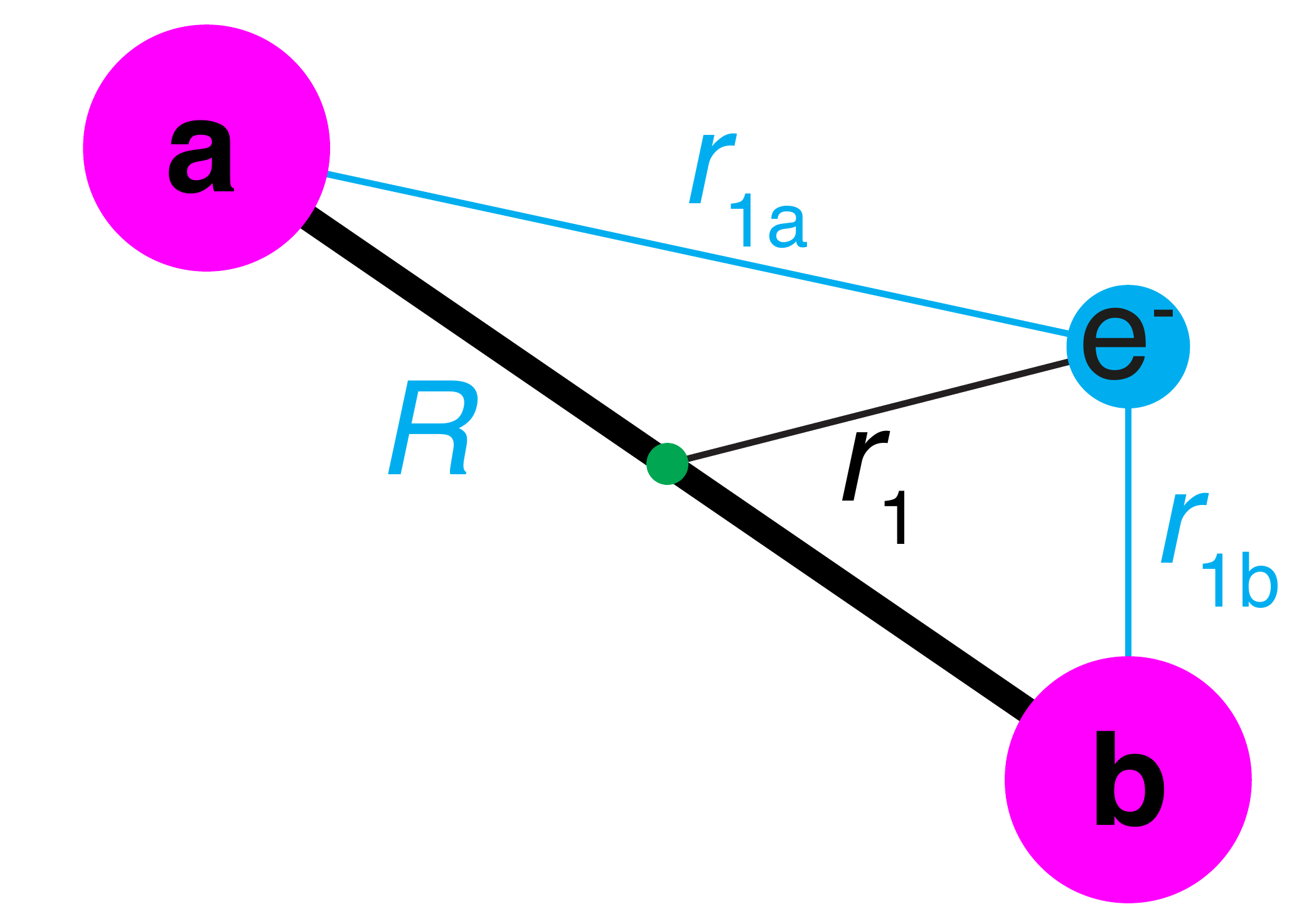}
	\caption{Relevant distances in the HD$^+$ molecular ion. The small blue dot designates the geometric centre of the nuclei. \label{fig0}}
\end{figure}

HD$^+$ is the simplest molecular system that possesses a permanent electric dipole moment. The nonvanishing electric dipole moment results from a breakdown of the Born-Oppenheimer approximation that can be attributed to the displacement of the centre-of-mass relative to the geometric centre of the nuclei. The rovibronic molecular Hamiltonian can be expressed as \cite{wolniewicz80a,carrington84a} 
\begin{equation}\label{H2+_hamiltonian}
\begin{split}
\mathcal{H_\text{rve}} &= \underbrace{-\frac{\nabla^2_1}{2}  + \frac{1}{R} - \frac{1}{r_{1\text{a}}} - \frac{1}{r_{1\text{b}}} }_{H_{\rm cn}} \\
& \underbrace{-\frac{\nabla^2_{R\Theta\Phi} }{2\mu}}_{H'_1}   \underbrace{-\frac{\nabla^2_1 }{8\mu} }_{H'_2} \underbrace{-\frac{\nabla_{R\Theta\Phi}\cdot \nabla_1  }{2\mu_\alpha}}_{H'_3}. 	
\end{split}
\end{equation}
In Eq.~(\ref{H2+_hamiltonian}), $H_{\rm cn}$ is the \emph{clamped-nuclei} Hamiltonian, $(R\Theta\Phi)$ specify the nuclear spatial arrangement in the laboratory-fixed frame, and 
\begin{equation}
\mu = \frac{m_\text{a}m_\text{b}}{m_\text{a} + m_\text{b}}
\end{equation}
and
\begin{equation}
\mu_\alpha = \frac{m_\text{a}m_\text{b}}{m_\text{a}-m_\text{b}},
\end{equation}
where $m_\text{a}$ and $m_\text{b}$ are the masses of nuclei a and b, respectively. All other symbols have their usual meaning and the labels of the particles are given in Fig.~\ref{fig0}.

The Hamiltonian of HD differs from that of the homonuclear species H$_2$ and D$_2$ only through the additional term $H'_3$, which vanishes in homonuclear molecules and is the only term in Eq.~(\ref{H2+_hamiltonian}) that is not invariant to the symmetry operations of the D$_{\infty {\rm h}}$ point group. Specifically, $H'_3$ is not invariant with respect to the permutation of the nuclei a and b and has C$_{\infty {\rm v}}$ symmetry. This term causes a mixing of states of $g$ and $u$ symmetry (see, e.g., Refs.~\cite{wolniewicz80a,carrington84a}) and couples ionisation channels differing in core rotational-angular-momentum quantum number $N^+$ and electron-orbital-angular-momentum quantum number $l$ by $\pm 1$ \cite{merkt96a,sprecher14c}. Because of the mass ratio of H to D (and T), the nonadiabatic {\it g/u}-symmetry breaking in HD (and HT) is stronger than in any other isotopically substituted homonuclear diatomic molecule. 

The rovibronic {\it g/u}-symmetry-breaking interaction is very small in low rovibrational states, but becomes important in high vibrational states near the dissociation thresholds of HD$^+$, H$^+$ + D(1s) and D$^+$ + H(1s), that are separated by the different ionisation energies of the atoms, i.e., by 29.843~cm$^{-1}$. These two dissociation thresholds can be associated with the lower (X$^+$) and upper (A$^+$) electronic states, respectively, and the theoretical description is similar to that introduced for the hyperfine-induced {\it g/u} mixing in H$_2^+$ presented in Ref.~\cite{beyer18d}, although the rovibronic {\it g/u} mixing effect in HD$^+$ is approximately 900 times stronger.

The nonzero electric dipole moment of HD$^+$ allows the measurement of rovibrational transitions in the infrared and microwave region and a variety of such transitions have been observed \cite{wing76a,carrington81a,carrington83a,carrington85a,carrington91a,carrington92a,alighanbari20a,patra20a}.
In particular, Carrington and coworkers have studied the hyperfine structure of rovibrational levels of HD$^+$ in the vicinity of the first dissociation threshold and the effects of the asymmetric charge distribution in the centre-of-mass frame \cite{carrington91a}.

Above the first dissociation threshold of HD$^+$, the {\it g/u}-mixing term $H^\prime_3$ in the molecular Hamiltonian~(\ref{H2+_hamiltonian}) is also responsible for electronic predissociation (type I predissociation in Herzberg's classification \cite{herzberg89a}). The rovibrational levels of the A$^+$ state cannot predissociate in the cases of H$_2^+$ and D$_2^+$ because they are located below the dissociation threshold. In HD$^+$, they are all located above the H$^+$ + D(1s) dissociation threshold and are coupled to the H$^+$ + D(1s) continuum, which reduces the lifetimes to the picosecond range.
Although the decay through electronic predissociation is allowed above the lowest dissociation threshold, it does not always contribute significantly, so that some resonances are better described by elastic-scattering resonances i.e., shape and orbiting resonances, as discussed, {\it e.g.}, in Refs.~\cite{davis78a,beyer16a,beyer18b}.

All bound rovibronic states of HD$^+$ associated with the X$^+$ state were calculated by Moss using a transformed Hamiltonian in combination with the artificial-channels scattering method \cite{moss93b}. With this method, the effect of the {\it g/u} mixing rovibronic term is removed from the Hamiltonian using a unitary transformation that results in effective nuclear charges and effective reduced masses in the electronic Hamiltonian.
For the lowest vibrational states of HD$^+$, variational calculations \cite{moss89a} and pre-Born-Oppenheimer variational results including relativistic and QED corrections up to high order \cite{korobov04a,korobov06a,korobov14a,korobov21a} were reported.

Calculations, based on a variation-perturbation approach first used in HD$^+$ by Wolniewicz and Orlikowski \cite{wolniewicz80a,wolniewicz91a, orlikowski94a}, using hyperspherical coordinates \cite{igarashi99a} and a modified adiabatic Hamiltonian \cite{esry99a}, were also reported (see Ref.~\cite{leach95a} for a review), but included more restrictive approximations than the calculations of Moss \cite{moss93b}. However, these calculations could be extended to the energy region above the first dissociation threshold H$^+$ + D(1s) and predicted the positions and widths of the quasibound states of HD$^+$ \cite{davis78a,wolniewicz91a, orlikowski94a,igarashi99a,esry00a}.

Several quasibound states (shape resonances) of the X$^+$ state with rotational quantum numbers $N^+$ larger than 16 were observed by Carrington and coworkers \cite{carrington88a}. Their experimental method restricted the detection to states with lifetimes longer than a few nanoseconds so that only resonances trapped on the low-$R$ side of large centrifugal barriers could be detected. Concerning the metastable levels of the A$^+$ state, Leach and Moss commented in their review article that ,,there seems to be no immediate prospect to experimentally observe these short-lived levels of the first excited electronic state'' \cite{leach95a}.

We present PFI-ZEKE photoelectron spectra of HD near the H$^+$ + D and H + D$^+$ dissociative-ionisation thresholds recorded via the H$\bar{\rm H}$~$^1\Sigma_g^+\ (v=11-13)$ and B$^{\prime\prime}\bar{\rm B}$~$^1\Sigma_u^+\ (v=11-13)$ intermediate states. These spectra reveal sharp transitions to the highest bound states of HD$^+$ with $v^+$ between 16 and 21 and $N^+=0-10$ as well as the onset of the dissociation continua. The effects of increasing {\it g/u} mixing with increasing $v^+$ on the relative intensities in these spectra are discussed. By detecting the H$^+$ and D$^+$ fragments using mass-analysed threshold-ionisation (MATI) spectroscopy, we also observed the metastable rovibrational levels of the A$^+$ state for the first time. These levels are located between the two dissociation thresholds and are Feshbach resonances subject to fast electronic predissociation. We also observed three other resonances that are attributed to the X$^+(21,4)$, $(20,7)$ and A$^+$(0,4) resonances, the first two of which are located below the second dissociation threshold D$^+$ + H(1s).

\section{Experimental setup and procedure}\label{exp_setup}
\begin{figure}\centering
	\includegraphics[width=0.9\textwidth]{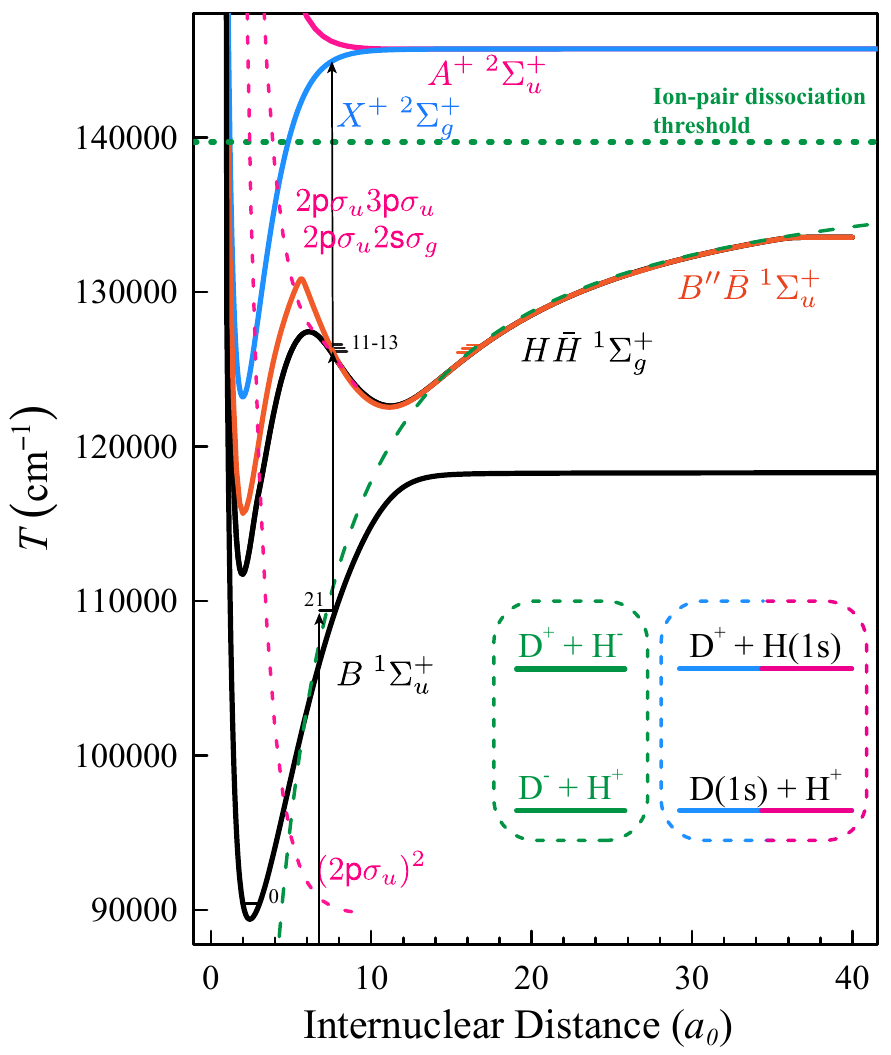}
	\caption{Potential-energy functions of the B~$^1\Sigma_u^+$, H$\bar{\rm H}$~$^1\Sigma_g^+$ and B$^{\prime\prime}\bar{\rm B}$~$^1\Sigma_u^+$ states of molecular hydrogen and of the X$^+$~$^2\Sigma_g^+$ and A$^+$~$^2\Sigma_u^+$ states of the molecular-hydrogen cation. The dashed lines represent schematically the diabatic potential-energy functions of the repulsive states with configurations $(2{\rm p}\sigma_u)^2$, $(2{\rm p}\sigma_u)(3{\rm p}\sigma_u^+)$ and $(2{\rm p}\sigma_u)(2{\rm s}\sigma_g)$ (magenta) and of the H$^+$D$^-$/H$^-$D$^+$ ion-pair states (green). The horizontal dotted green line gives the position the ion-pair-dissociation thresholds. The left inset indicates the two ion-pair dissociation thresholds D$^+$ + H$^-$ and D$^-$ + H$^+$ of HD, and the right inset the two dissociation thresholds D$^+$ + H(1s) and D(1s) + H$^+$ of HD$^+$. The blue and magenta colours highlight the mixed characters in the correlation to the X$^+$ and A$^+$ states. The vertical arrows indicate the photoexcitation sequence used to access the region of the dissociative-ionisation threshold from the X~$^1\Sigma_g^+$ ground state of HD. 
		\label{Fig0_pot}}
\end{figure}

The experiments relied on the use of PFI-ZEKE photoelectron spectroscopy \cite{muellerdethlefs91b} in combination with a resonant three-photon-excitation scheme from the ground X~$^1\Sigma_g^+~(v=0)$ state of HD. The region around the first dissociative-ionisation threshold of HD, which is located 146084.55541(37)~cm$^{-1}$ above the X~$^1\Sigma_g^+\ (v=0,N=0)$ ground state of HD  \cite{sprecher10a,moss93b}, was accessed using the resonant three-photon excitation sequence
\begin{align}
\label{CRHD:eq:excitation}
\text{X}~^1\Sigma_g^+~(0,0-3) &\xrightarrow{\rm VUV} \text{B}~^1\Sigma_u^+~(20-21,1-4) \notag \\ 
&\xrightarrow{\rm VIS} \bar{\rm H}~^1\Sigma_g^+~(11-13,0-5)\quad\text{or}\quad \bar{\rm B}~^1\Sigma_u^+~(11-13,1-4)\quad \notag\\ 
&\xrightarrow{\rm UV} \text{X}^+~^2\Sigma_g^+~(16-21,0-10) \quad\text{and}\quad \text{A}^+~^2\Sigma_u^+~(0,0-4)
\end{align}
via excited vibrational levels of the B~$^1\Sigma_u^+$ and either the $\bar{\rm H}~^1\Sigma_g^+$ or the $\bar{\rm B}~^1\Sigma_u^+$ intermediate states, see Fig.~\ref{Fig0_pot}. The same laser system, consisting of a broadly tunable VUV laser used to drive the B-X transition, a visible laser to induce the H$\bar{\rm H}$-B and B$^{\prime\prime}\bar{\rm B}$-B transitions, and a UV laser to access the dissociative-ionisation threshold, was already used to study the highest bound states and shape resonances of H$_2^+$ and D$_2^+$ and is described in Refs.~\cite{beyer16a,beyer16b,beyer17a,beyer18b}.  

The {\it g/u}-symmetry breaking in HD allowed the direct excitation, from the ${\rm B}~^1\Sigma_u^+~(20-21,1-4)$ intermediate levels, of rovibrational levels of the B$^{\prime\prime}\bar{\rm B}$ $^1\Sigma_u^+ $state \cite{reinhold98a,reinhold99a}, which has (1s$\sigma_g$)3p$\sigma_u$ character in the inner well and (2p$\sigma_u$)2s$\sigma_g$ and ion-pair character in the outer well.
Because of the ion-pair character of the outer-well states of the H$\bar{\rm H}$ and B$^{\prime\prime}\bar{\rm B}$ state, their potential-energy curves are almost identical beyond $8\,a_0$. Consequently, the {\it g/u} mixing is substantial and some levels of the B$^{\prime\prime}\bar{\rm B}$ state can be excited from the B state almost as efficiently as the corresponding levels of the H$\bar{\rm H}$ state despite the fact that their nominal \emph{ungerade} character makes the excitation forbidden in zero order.
\begin{figure}\centering
	\includegraphics[width=0.9\textwidth]{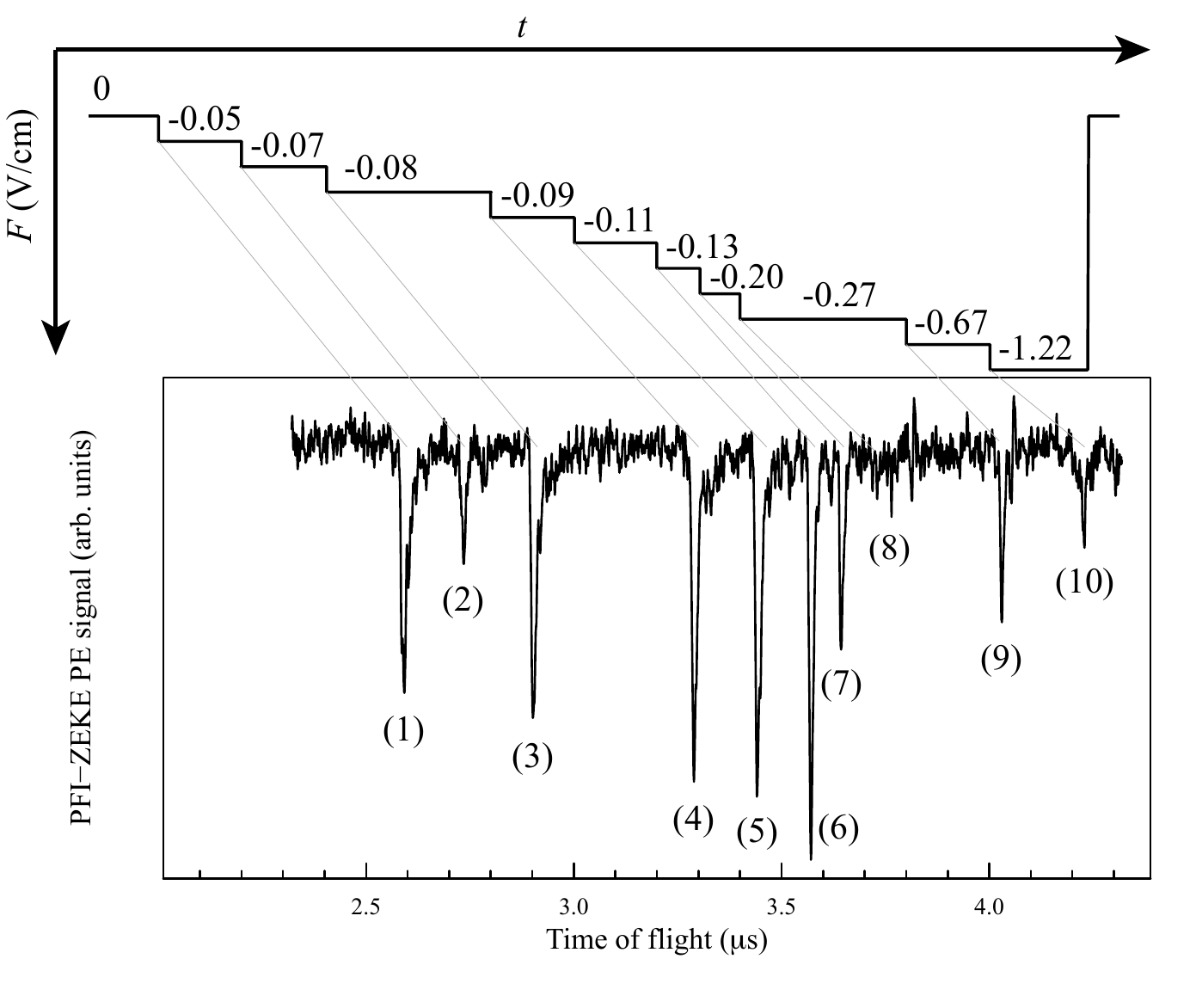}
	\caption{Upper Panel: Electric-field pulse sequence used to record the PFI-ZEKE photoelectron spectrum of HD in the vicinity of the dissociative-ionisation thresholds of HD. Lower panel: Electron time-of-flight spectrum obtained by collecting the electrons generated by the pulsed-field-ionisation sequence. The peak labels (1)-(10) designate the field pulses that led to the field ionisation of the corresponding electrons, as indicated by the grey diagonal lines. 
		\label{ZEKE_sequence}}
\end{figure}

The PFI-ZEKE photoelectron spectra were recorded by monitoring the yield of electrons generated by delayed pulsed field ionisation as a function of the wave number of the UV laser. The multipulse electric-field sequence used for field ionisation is displayed in the upper panel of Fig.~\ref{ZEKE_sequence} and consists of 10 electric-field steps with field strength increasing from $-0.05$ to $-1.22$~V/cm. The electric fields also extracted the electrons toward a microchannel-plate (MCP) detector located a the end of a flight tube.
The lengths of the individual pulses were chosen so as to be able to unambiguously relate each electron time-of-flight peak to the corresponding electric-field step in the sequence. Their strengths were selected to achieve a high resolution (about 0.15~cm$^{-1}$) in the spectra generated by pulses (2)-(5) and a high signal-to-noise ratio in the spectra recorded with the last pulses of the sequence. The lower part of Fig.~\ref{ZEKE_sequence}  presents the time-of-flight spectrum of the electrons generated by the pulse sequence when carrying out the photoexcitation to a spectral position above the dissociative-ionisation thresholds, which makes it possible to generate a field-ionisation signal simultaneously with all field pulses. The labels (1) to (10) designate the groups of electrons generated by the 10 pulses of the sequence. The PFI-ZEKE photoelectron spectra presented in the next section correspond to the electron signals generated by the field pulses (2)-(5). The positions of the ionisation thresholds were corrected for the field-induced shifts, which were determined from calculations of the field-ionisation rates of the Rydberg-Stark states following the procedure described in Ref.~\cite{hollenstein01a}. The field-corrected spectra were then added to improve the signal-to-noise ratio.

MATI spectra were also recorded in the vicinity of the dissociative-ionisation thresholds using a sequence of two electric-field pulses. The first, weak ($-70$~mV/cm) discrimination pulse served the purpose of sweeping prompt ions out of the photoexcitation volume whereas the second, stronger (800 mV/cm) pulse was used to field ionise high Rydberg states located below the bound and continuum states of HD$^+$ and extract the H$^+$, D$^+$ and HD$^+$ ions toward the MCP detector. Below the D(1s) + H$^+$ dissociation threshold, only HD$^+$ ions corresponding to weakly bound vibrational levels of the X$^+$ and A$^+$ states can be observed in the MATI spectra for energetic reasons. H$^+$ and D$^+$ ions can only be observed above the D(1s) + H$^+$  and the D$^+$ + H(1s) dissociation thresholds, respectively.

The UV wave number was calibrated using a wavemeter (absolute accuracy of 0.02~cm$^{-1}$, relative accuracy of 0.015~cm$^{-1}$) and the line centres were determined using a Poisson-weighted nonlinear fit of a Gaussian line-shape model assuming a constant line width determined by the field-ionisation sequence. The relative positions of the rovibronic levels of HD$^+$ were extracted in a weighted linear least-squares fit from a redundant network of more than 350 transitions connecting different rovibrational levels of the $\bar{\rm H}$ and $\bar{\rm B}$ states to the rovibronic ionisation thresholds. The absolute term values of the HD$^+$ levels with respect to the X~$^1\Sigma_g^+(0,0)$ ground state of HD  were obtained by adding the calibrated UV wave numbers to the term values of the $\bar{\rm H}~^1\Sigma_g^+$ or $\bar{\rm B}~^1\Sigma_u^+$ levels reported by Reinhold {\it et al.} \cite{reinhold99a} and compensating for the shifts of the ionisation thresholds induced by the pulsed field ionisation~\cite{hollenstein01a}. 
This procedure resulted in typical absolute and relative uncertainties of 0.11~cm$^{-1}$ and 0.02~cm$^{-1}$, respectively, for the rovibronic levels of HD$^+$. 

\section{Experimental results}\label{results}

\subsection{ Structure and dynamics of the H$\bar{H}(v=11-13)$ and B$^{\prime\prime}\bar{B}(v=11-13)$ intermediate states} \label{sec:HB_dynamics}

\begin{figure}\centering
	\includegraphics[width=0.8\textheight,angle=90]{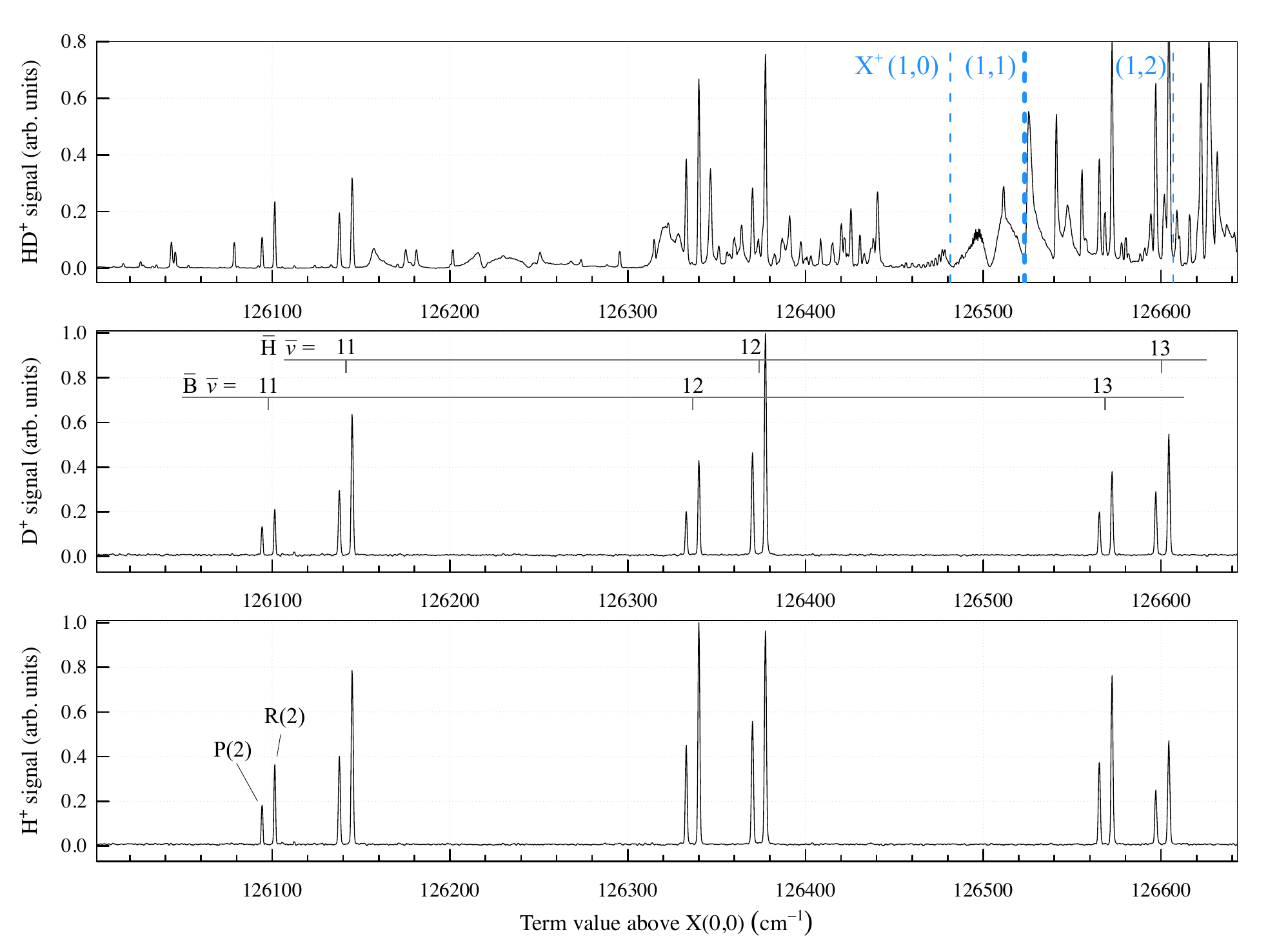}
	\caption{Photoionisation spectrum of HD in the vicinity of the X$^+(v^+=1,N^+=0-2)$ levels of HD$^+$ (indicated by dashed vertical lines) recorded via the B (20,2) level. The HD$^+$, D$^+$ and H$^+$ ion signals are shown in the top, middle and bottom panels, respectively. The pairs of lines correspond to the P(2) and R(2) transitions to the $\bar{v}=11-13$ vibrational states of the $\bar{\rm H}~^1\Sigma_g^+$ and $\bar{\rm B}~^1\Sigma_u^+$ states. See text for details.
		\label{CRHD:fig:spec_HH_BB}}
\end{figure}

The (1+1') resonant two-photon ionisation spectrum of HD in the vicinity of the X$^+(v^+=1,N^+=0-2)$ levels of HD$^+$ recorded via the B $^1\Sigma_u^+$ (20,2) intermediate level is displayed in Fig.~\ref{CRHD:fig:spec_HH_BB}. An electric field of 260~V/cm was applied 100~ns after the VUV and VIS excitation to (i) field ionise molecular Rydberg states of HD and (ii) extract all ions toward the MCP detector. HD$^+$, D$^+$ and H$^+$ ion signals could be observed separately using the TOF spectrometer and the corresponding spectra are displayed in the top, middle and bottom panels, respectively.

The dominant contributions to the HD$^+$ signal in the top panel of Fig.~\ref{CRHD:fig:spec_HH_BB} are from molecular Rydberg states, even though these states have low principal quantum numbers and cannot be field ionised by the 260~V/cm electric-field pulse. The HD$^+$ signal in this case is caused by (field-induced) autoionisation. 
In this spectral region, the Rydberg states excited from the B state, which has (1s$\sigma_g$)(2p$\sigma_u$) character at short range, are s and d Rydberg states converging on the X$^+(1,1)$ level of HD$^+$. Although HD does not have an inversion centre, the {\it g/u} symmetry of the electronic wavefunctions is a good symmetry within the Born-Oppenheimer approximation and is only broken by nonadiabatic interactions, as explained in the introduction. Consequently, the {\it g/u} symmetry remains useful as approximate symmetry in HD and can be used to explain intensity patterns. Moreover, HD, unlike H$_2$ and D$_2$, is not subject to restrictions imposed by the Pauli principle. Starting from a rotational state with odd rotational angular momentum in the ground state (e.g., X$(0,1)$), we expect the strongest autoionising Rydberg series to be s and d Rydberg series converging on the X$^+(1,1)$ state of HD$^+$ and such series can indeed be seen above 126460~cm$^{-1}$ in the top panel of Fig.~\ref{CRHD:fig:spec_HH_BB}. Some intensity is also observed for the $n$d$3_2\,(v^+=1)$ Rydberg series, which gives rise to the broad rotational-autoionisation resonances with Fano-type lineshapes above the X$^+(1,0)$ ionisation threshold. Rydberg series converging on higher vibrational levels of HD$^+$ also contribute to the spectrum depicted in the top panel of Fig.~\ref{CRHD:fig:spec_HH_BB}.

The rovibrational levels of the $\bar{\rm H}~^1\Sigma_g^+$ and $\bar{\rm B}~^1\Sigma_u^+$ outer-well states that we wanted to use as intermediate levels to access the region of the dissociative-ionisation thresholds of HD are difficult to recognize in the top panel of Fig.~\ref{CRHD:fig:spec_HH_BB} but they are clearly visible in the middle and bottom panels, which display the D$^+$ and H$^+$ ion signals, respectively, observed in the region of the $v=11$-13 vibrational levels.
The rovibrational states in the $\bar{\rm H}$ and $\bar{\rm B}$ outer wells of the H$\bar{\rm H}$ and B$^{\prime\prime}\bar{\rm B}$ states are known to predissociate to the H(2$l$) + D(1s) and H(1s) + D(2$l$) dissociation continua of lower electronic states of $\Sigma_{g/u}$ and $\Pi_{g/u}$ symmetry \cite{reinhold00a}.
The H$^+$ and D$^+$ ions observed in the middle and bottom panels of Fig.~\ref{CRHD:fig:spec_HH_BB}, respectively, result from predissociation of the rovibrational levels of the $\bar{\rm H}$ and $\bar{\rm B}$ states forming H and D atoms in $n=2$ Rydberg states. These Rydberg states are then ionised by the VIS laser pulse in a multiphoton process to form H$^+$ and D$^+$ ions. Reinhold \emph{et al.} found lifetimes shorter than 5~ns for the low vibrational levels of the H$\bar{\rm H}$ state of HD compared to lifetimes of the order of 20~ns in H$_2$ and 50~ns in D$_2$ \cite{reinhold00a}. They concluded that the radiative decay is stronger in HD because of the larger number of dipole-allowed transitions to lower states originating from {\it g/u}-symmetry mixing. Our observation of strong HD$^+$, H$^+$ and D$^+$ signals following excitation to the $\bar{\rm H}$ and $\bar{\rm B}$ states indicates that autoionisation and predissociation are also important decay channels for vibrational levels in the range $v=11$-13 , the {\it g/u}-symmetry mixing allowing nonadiabatic interactions with additional continua compared to H$_2$ and D$_2$. 

The spectra depicted in the middle and bottom panels of Fig.~\ref{CRHD:fig:spec_HH_BB} are dominated by the P(2) and R(2) lines of the $\bar{\rm H}$(11-13) -- B(20) and $\bar{\rm B}$(11-13) -- B(20) bands. The rovibrational levels of the H$\bar{\rm H}$ and B$^{\prime\prime}\bar{\rm B}$ states are closely spaced, as expected from the large internuclear distances of the outer-well states. The degeneracy of the potential energies of the  $\bar{\rm H}$ and $\bar{\rm B}$ states at long range (see Fig.~\ref{Fig0_pot}) implies a close proximity of their vibrational levels and a significant mixing of the {\it g/u} character in HD caused by nonadiabatic interactions. The spectral lines associated with each value of $v$ thus form two pairs, the lower and higher pairs corresponding to exitation of the $\bar{\rm B}$ and $\bar{\rm H}$ states, respectively.  The $\bar{\rm H}$ and $\bar{\rm B}$ states have ion-pair character at large internuclear distance and can be represented as positive (\emph{gerade}) and negative (\emph{ungerade}) superposition of the H$^+$D$^-$(1s)$^2$ and H$^-$(1s)$^2$D$^+$ ion-pair configurations. The {\it g/u} mixing results in a superposition of these \emph{gerade} and \emph{ungerade} ion-pair configurations, which results in the (at least partial) localisation of the electrons on either the proton or the deuteron. Based on the difference in the ionisation energies and electron affinities of atomic hydrogen and deuterium, we expect the lower-lying pair of states, i.e., corresponding to the $\bar{\rm B}$ state, to be correlated with the H$^+$D$^-$ ion-pair configuration and the higher-lying pair of states (corresponding to the $\bar{\rm H}$ state) to be correlated with the H$^-$D$^+$ ion-pair configuration.

The relative intensities of the transitions to the $v=11$-13 levels of the $\bar{\rm H}$ and $\bar{\rm B}$ states observed in the three spectra depicted in Fig.~\ref{CRHD:fig:spec_HH_BB} is determined by the complex interplay between the transition moments from the B(20) intermediate level and the competition between radiative decay and decay by predissociation and autoionisation. The B state of HD around $v=20$ has dominant {\it u} character so that the absorption intensities are proportional to the {\it g} character of the $\bar{\rm H}$(11-13) and $\bar{\rm B}$(11-13) levels. These characters can be derived from the analysis of these states presented in Refs.~\cite{wolniewicz98a,reinhold99a} to be 0.86, 0.81, and 0.72 for the $v=11$, 12 and 13 levels of the $\bar{\rm H}$ state and 0.14, 0.19 and 0.28 for the $v=11$, 12 and 13 levels of the $\bar{\rm B}$ state, with almost no dependence on the rotational quantum number. These characters explain why the lines associated with the excitation of the $\bar{\rm H}$ vibrational levels are overall stronger than the lines associated with the corresponding $\bar{\rm B}$ vibrational levels. However, the observed intensity ratios are less pronounced than would be expected on the basis of the {\it g} characters, presumably because of differences in the radiative decay of the $\bar{\rm H}$ and $\bar{\rm B}$ states \cite{reinhold00a}. 

The intensities of the transitions to the $v=11$-13 states of the $\bar{\rm H}$ and $\bar{\rm B}$ states observed in the HD$^+$ channel (top panel of Fig.~\ref{CRHD:fig:spec_HH_BB}) increase rapidly with increasing $v$ value. This increase can be readily understood by considering the potential-energy curves depicted in Fig.~\ref{Fig0_pot}. In the spectral region under investigation, autoionisation must take place to the ionisation continua associated with the X$^+(v^+=0,1)$ vibronic states of HD$^+$ for energetic reasons. The autoionisation rate is given by the overlap of the vibrational wavefunctions of the $\bar{\rm H}$ and $\bar{\rm B}$ states with the X$^+(v^+=0,1)$ vibrational wavefunction and, consequently, by the probability of tunnelling through the barriers separating the two potential wells of the H$\bar{\rm H}$ and B$^{\prime\prime}\bar{\rm B}$ states. This probability increases with increasing $v$ value.

The relative intensities of the transitions to the $\bar{\rm H}$ and $\bar{\rm B}$ states in the spectra depicted in the middle and lower panels of Fig.~\ref{CRHD:fig:spec_HH_BB} indicate a slight preference for the $\bar{\rm B}$ levels to predissociate to the H(1s) + D($n=2$) continua and for the $\bar{\rm H}$ levels to predissociate to the D(1s) + H($n=2$) continua. This behaviour reflects the facts that the $\bar{\rm B}$ state correlates with the H$^+$D$^-$ ion-pair configuration and that the higher-lying $\bar{\rm H}$ state is correlated with the H$^-$D$^+$ ion-pair configuration, as mentioned above.
Predissociation is a vibronic effect and it is unlikely that both electrons are affected. The dominant predissociation processes are therefore expected to be the one-electron charge-transfer (CT) processes
\begin{equation}
{\rm H}^+{\rm D}^-(1{\rm s}^2) \xrightarrow{\rm CT\ predissociation} {\rm H}(2l) + {\rm D}(1{\rm s}) \label{CRHD:eq:pred_B}
\end{equation}
\begin{equation}
{\rm H}^-(1 {\rm s}^2){\rm D}^+ \xrightarrow{\rm CT\ predissociation} {\rm H}(1{\rm s}) + {\rm D}(2l) \label{CRHD:eq:pred_H}, 
\end{equation}
in accord with the trends in relative intensities observed in the lower two panels of Fig.~\ref{CRHD:fig:spec_HH_BB}.

\subsection{Bound levels of HD$^+$ just below the dissociative-ionisation thresholds of HD} \label{sec:bound}

\begin{figure}\centering
	\includegraphics[height=0.8\textheight,angle=0]{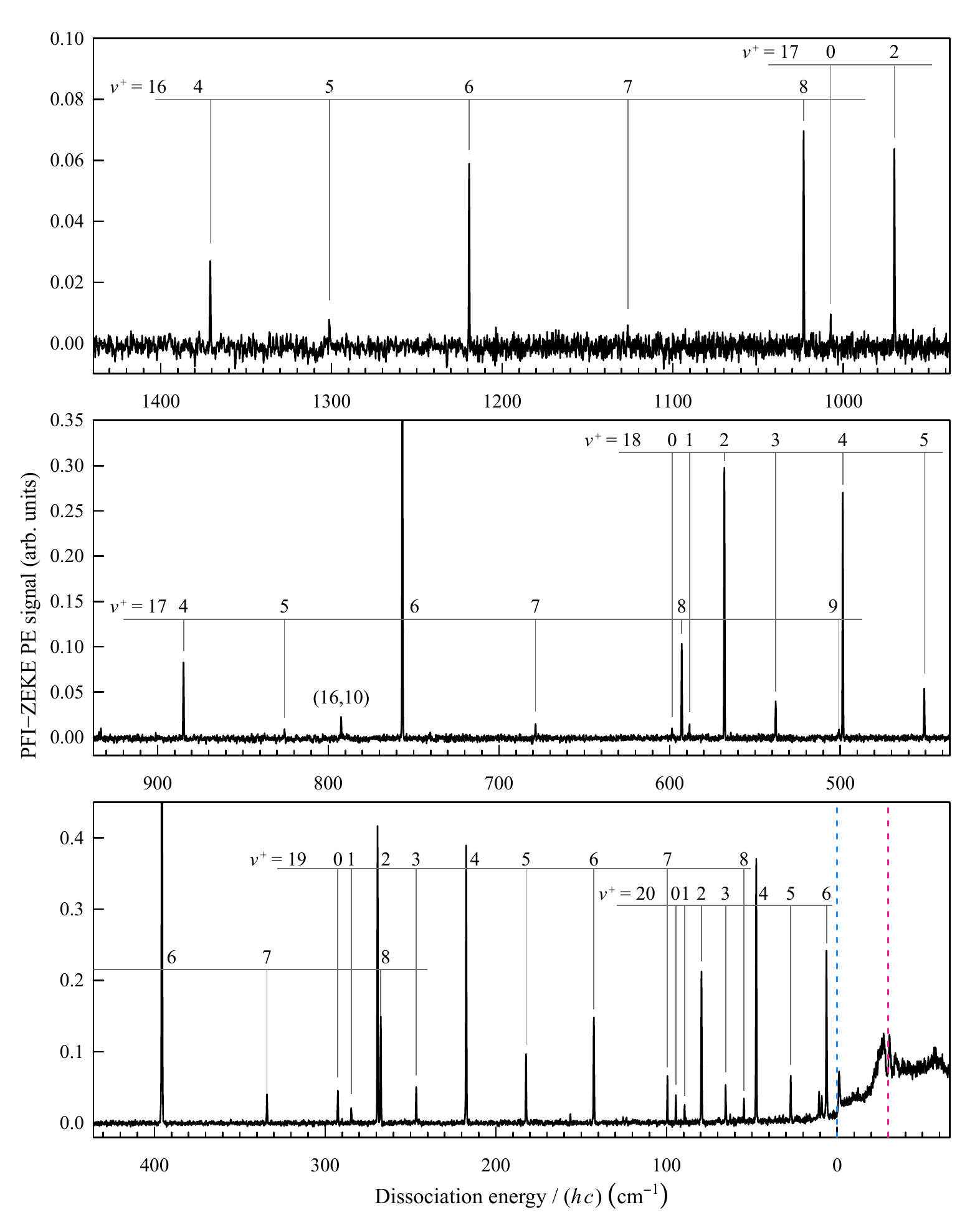}
	\caption{PFI-ZEKE photoelectron spectrum of HD recorded near the dissociative-ionisation threshold from the $\bar{\rm H}~^1\Sigma_g^+~(11,4)$ intermediate level. The H$^+$ + D(1s) and D$^+$ + H(1s) dissociation thresholds are indicated by dashed blue and red lines, respectively. \label{CRHD:fig:spec_HH_zeke}}
\end{figure}

The PFI-ZEKE PE spectrum of HD recorded via the $\bar{\rm H}~^1\Sigma_g^+~(11,4)$ intermediate level is depicted in Fig.~\ref{CRHD:fig:spec_HH_zeke}. The spectrum consists of transitions to bound levels of HD$^+$ that can be grouped in rotational progressions associated with vibrational levels with $v^+$ between 16 and 21 and $N^+$ between 0 and 10.
In H$_2$ or D$_2$, the conservation of total parity and nuclear-spin symmetry implies that photoionisation to the X$^+$ $^2\Sigma_g^+$ state of H$_2^+$ of D$_2^+$ from a \emph{gerade} intermediate state with even (odd) rotational-angular-momentum quantum numbers $N$ exclusively yields ions in rotational levels of even (odd) $N^+$ values \cite{xie90a,signorell97c}. This rotational photoionisation selection rule can be expressed as $\Delta N= N^+-N=0, \pm2, \pm4, \ldots$. If the photoionisation is from an \emph{ungerade} intermediate state, the selection rule is $\Delta N= \pm1, \pm3, \ldots$. In HD, there are no restrictions imposed by the conservation of nuclear-spin symmetry and one has to consider the effects of {\it g/u} mixing both in the intermediate and the ionic states.  Figure~\ref{CRHD:fig:spec_HH_zeke} shows that both even- and odd-$N^+$ levels of HD$^+$ are accessed from the $\bar{\rm H}$(11,4) intermediate level but that the transitions to the even-$N^+$ levels are much more intense than the transitions to the odd-$N^+$ levels. In the rotational progressions associated with $v^+=16$ and 17, the intensity ratio of the transitions to even and odd $N^+$ values roughly reflects the 0.86:0.14 ratio of {\it g} and {\it u} character of the $\bar{\rm H}(11)$ intermediate state. This observation suggests that the $v^+=16$ and 17 vibrational levels of the X$^+$ $^2\Sigma_g^+$ state of HD$^+$ have almost pure {\it g} character. With increasing value of $v^+$, however, the intensities of transitions to states of even and odd values of $N^+$ become more and more similar, indicating that {\it g/u} mixing in the HD$^+$ ion becomes significant beyond $v^+=18$.

\begin{figure}\centering
	\includegraphics[width=0.8\textheight,angle=90]{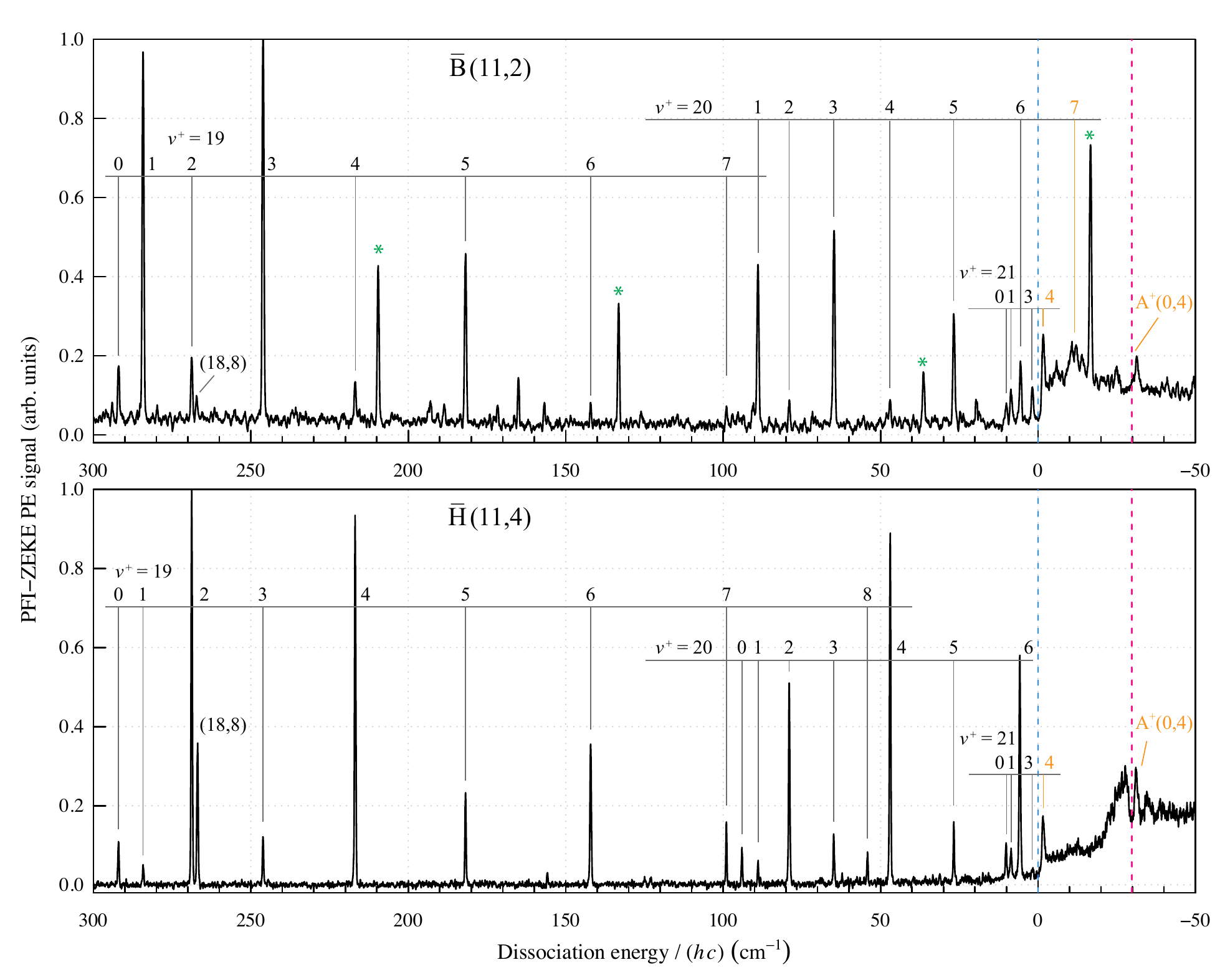}
	\caption{Comparison of the PFI-ZEKE photoelectron spectrum of HD near the dissociative-ionisation threshold recorded from the $\bar{\rm B}~^1\Sigma_g^+~(11,2)$ (upper panel) and $\bar{\rm H}~^1\Sigma_g^+~(11,4)$ (lower panel) levels. The H$^+$ + D(1s) and D$^+$ + H(1s) dissociation thresholds are indicated by dashed blue and red lines, respectively. Unassigned lines are marked with green asterisks. See text for details. \label{CRHD:fig:spec_HH_BB_zeke}}
\end{figure}

\begin{figure}\centering
	\includegraphics[width=0.8\textwidth]{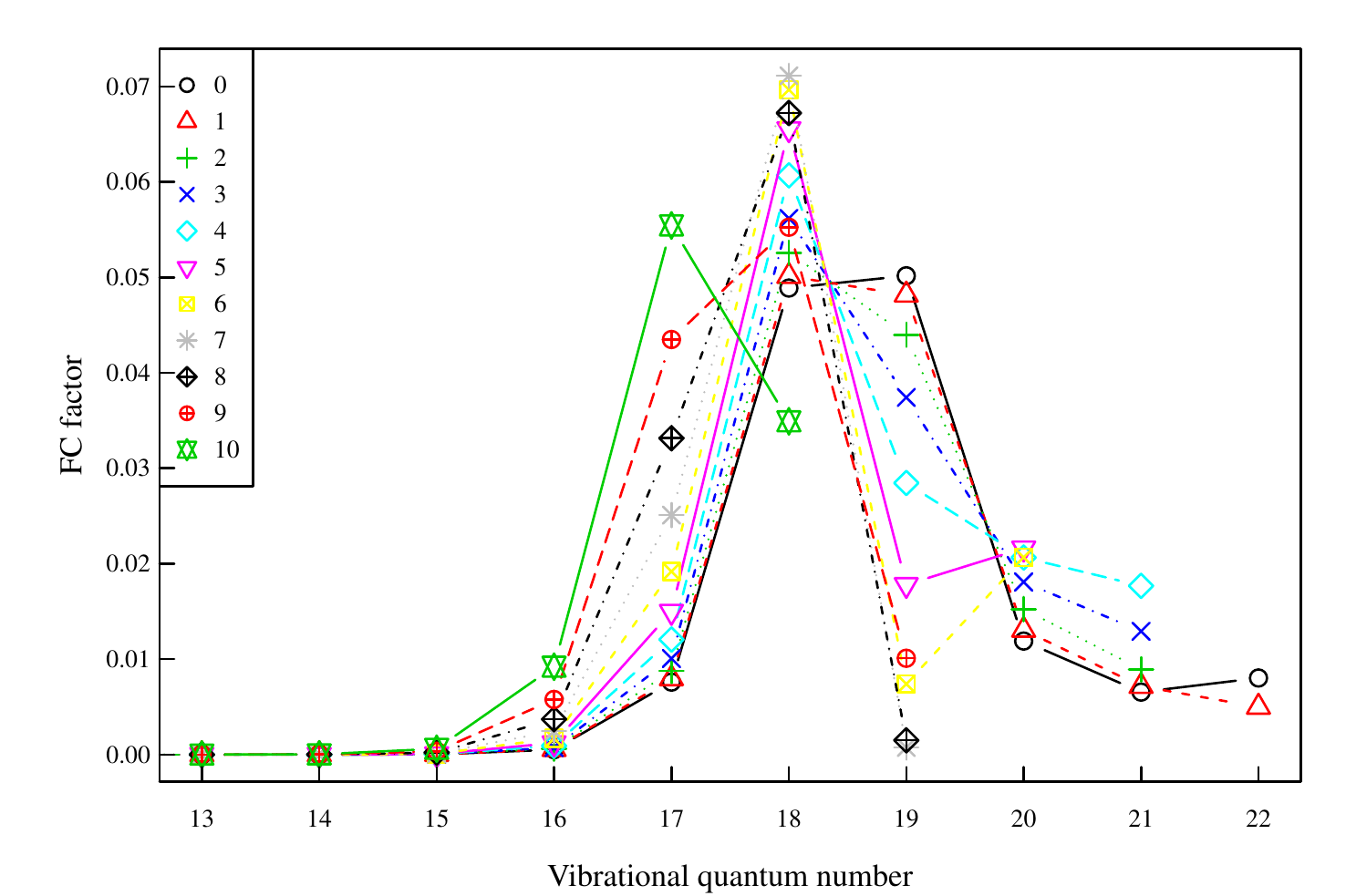}
	\caption{Franck-Condon factors for the excitation from the H$\bar{\rm H}$ $(v=11)$ state to selected rotational levels ($N^+$) of the X$^+$ state of HD$^+$. The lines connecting the data points are used to guide the eye. \label{CRHD:fig:FC_HH_X+_zeke}}
\end{figure}

The PFI-ZEKE PE spectra of HD recorded from the $\bar{\rm B}$(11,2) and $\bar{\rm H}$(11,4) intermediate states are compared in Fig.~\ref{CRHD:fig:spec_HH_BB_zeke}. Both intermediate states have even rotational quantum numbers, but opposite nominal {\it g/u} symmetry and thus display the opposite $N^+$-propensity rule.
To rule out that the Franck-Condon factors are the origin of the strong $N^+$-dependent intensity alternations in the spectra presented in Figs.~\ref{CRHD:fig:spec_HH_zeke} and~\ref{CRHD:fig:spec_HH_BB_zeke}, we calculated them in the adiabatic approximation and present them in Fig.~\ref{CRHD:fig:FC_HH_X+_zeke} (see discussion below concerning the adiabatic approximation). Within a rotational progression of a given vibrational state of HD$^+$, the Franck-Condon factors are found to either smoothly increase (e.g., for $v^+=17$) or smoothly decrease (e.g., for $v^+=19$), but no dependence on the even/odd nature of $N^+$ is observed.

The lines marked with asterisks in Fig.~\ref{CRHD:fig:spec_HH_BB_zeke} could not be assigned. They show the same field-ionisation shifts as the lines that could unambiguoulsy be assigned to transitions to rovibrational levels of the X$^+$ state, but do not correspond to the positions expected for transitions to these levels from the intermediate $\bar{\rm H}$ or $\bar{\rm B}$ states nor from the B state. Unlike the assigned lines, these lines can be eliminated by applying a prepulse in the pulsed-field ionisation sequence, and thus be distinguished from the transitions to the high-$v^+$ levels discussed here. Moreover, the additional lines appear at different absolute energies when other rovibrational levels of the intermediate state are used in the multiphoton excitation sequence [Eq.~\eqref{CRHD:eq:excitation}]. They may correspond to transitions from unknown levels which are populated through the radiative decay of the intermediate states.

The level positions, relative to the position of the X$^+$(20,4) level, of the highest bound states of HD$^+$ determined experimentally are listed in the fourth column of Table~\ref{CRHD:tab:level_eng}. Comparison with the theoretical dissociation energies including relativistic and radiative corrections reported by Moss \cite{moss93b} (fifth column) shows excellent agreement with our results, i.e., within the experimental uncertainties. The term values of all levels with respect to the X(0,0) ground state of HD are listed in the last column. Their absolute uncertainties are larger (about 0.11~cm$^{-1}$) because of the uncertainties in the field-induced shift of the ionisation threshold and in the term values of the intermediate levels (see Section~\ref{exp_setup}).

\begin{table}[t]
	\caption{Measured level positions (obs) of HD$^+$ with respect to the X$^+$(20,4) level. The columns labelled "calc" and "obs$-$calc" list the dissociation energies given by Moss \cite{moss93b} and their difference to our observed values. All values are in cm$^{-1}$. See text for details.}\label{CRHD:tab:level_eng}
		\begin{tabular}{l l l c r r}
			\hline
			      & $v^+$ & $N^+$ & \text{obs}    & calc~\cite{moss93b} & obs-calc \\ \hline
			$X^+$ & 16    & 4     & -1323.54(4)   & -1323.4819          & -0.058   \\
			$X^+$ & 16    & 5     & -1253.70(9)   & -1253.6991          & -0.001   \\
			$X^+$ & 16    & 6     & -1171.819(13) & -1171.8127          & -0.006   \\
			$X^+$ & 16    & 8     & -975.802(14)  & -975.7955           & -0.007   \\
			$X^+$ & 16    & 10    & -744.990(22)  & -744.9787           & -0.011   \\
			$X^+$ & 17    & 0     & -960.05(3)    & -960.0882           & 0.038    \\
			$X^+$ & 17    & 2     & -922.644(14)  & -922.6533           & 0.009    \\
			$X^+$ & 17    & 4     & -837.360(13)  & -837.3674           & 0.007    \\
			$X^+$ & 17    & 5     & -778.18(6)    & -778.2028           & 0.023    \\
			$X^+$ & 17    & 6     & -709.125(11)  & -709.1187           & -0.006   \\
			$X^+$ & 17    & 7     & -631.147(25)  & -631.1418           & -0.005   \\
			$X^+$ & 17    & 8     & -545.455(12)  & -545.4622           & 0.007    \\
			$X^+$ & 17    & 9     & -453.39(8)    & -453.4392           & 0.049    \\
			$X^+$ & 18    & 0     & -551.11(5)    & -551.1163           & 0.006    \\
			$X^+$ & 18    & 1     & -540.87(4)    & -540.8346           & -0.035   \\
			$X^+$ & 18    & 2     & -520.464(12)  & -520.4569           & -0.007   \\
			$X^+$ & 18    & 3     & -490.380(17)  & -490.3545           & -0.025   \\
			$X^+$ & 18    & 4     & -451.082(12)  & -451.0830           & 0.001    \\
			$X^+$ & 18    & 5     & -403.392(12)  & -403.3827           & -0.009   \\
			$X^+$ & 18    & 6     & -348.178(11)  & -348.1802           & 0.002    \\
			$X^+$ & 18    & 7     & -286.612(15)  & -286.5951           & -0.017   \\
			$X^+$ & 18    & 8     & -219.978(12)  & -219.9544           & -0.024   \\
			$X^+$ & 18    & 10    & -77.97(8)     & -78.0598            & 0.090    \\
			$X^+$ & 19    & 0     & -245.100(17)  & -245.0871           & -0.013   \\
			$X^+$ & 19    & 1     & -237.256(23)  & -237.2628           & 0.007    \\
			$X^+$ & 19    & 2     & -221.819(12)  & -221.8163           & -0.003   \\
			$X^+$ & 19    & 3     & -199.176(14)  & -199.1536           & -0.022   \\
			$X^+$ & 19    & 4     & -169.903(12)  & -169.8886           & -0.014   \\
			$X^+$ & 19    & 5     & -134.859(14)  & -134.8521           & -0.007   \\
			$X^+$ & 19    & 6     & -95.113(12)   & -95.1101            & -0.003   \\
			$X^+$ & 19    & 7     & -52.024(15)   & -52.0012            & -0.023   \\
			$X^+$ & 19    & 8     & -7.21(3)      & -7.2188             & 0.009    \\
			$X^+$ & 20    & 0     & -47.067(16)   & -47.0614            & -0.006   \\
			$X^+$ & 20    & 1     & -42.004(21)   & -41.9851            & -0.019   \\
			$X^+$ & 20    & 2     & -32.065(12)   & -32.0733            & 0.008    \\
			$X^+$ & 20    & 3     & -17.846(20)   & -17.8193            & -0.027   \\
			$X^+$ & 20    & 4     & 0$^a$         & 0$^a$               & 0$^a$    \\
			$X^+$ & 20    & 5     & 20.233(17)    & 20.2435             & -0.011   \\
			$X^+$ & 20    & 6     & 41.186(11)    & 41.1419             & 0.044    \\
			$X^+$ & 20    & 7     & 58.17(9)      & $\dots$             & $\dots$  \\
			$X^+$ & 21    & 0     & 36.798(16)    & 36.7796             & 0.018    \\
			$X^+$ & 21    & 1     & 38.468(16)    & 38.4424             & 0.026    \\
			$X^+$ & 21    & 2     & 41.374(22)    & 41.4426             & -0.069   \\
			$X^+$ & 21    & 3     & 45.17(3)      & 45.1564             & 0.014    \\
			$X^+$ & 21    & 4     & 48.647(22)    & $\dots$             & $\dots$  \\ \hline
			\multicolumn{6}{l}{\parbox[t]{9.5cm}{\footnotesize
	{$^a$}Reference level determined at be located 146037.73(18)~cm$^{-1}$ above the ground rovibronic state of HD.
	}}
		\end{tabular}

\end{table}

\begin{figure}
	\centering
	\includegraphics[width=1\columnwidth]{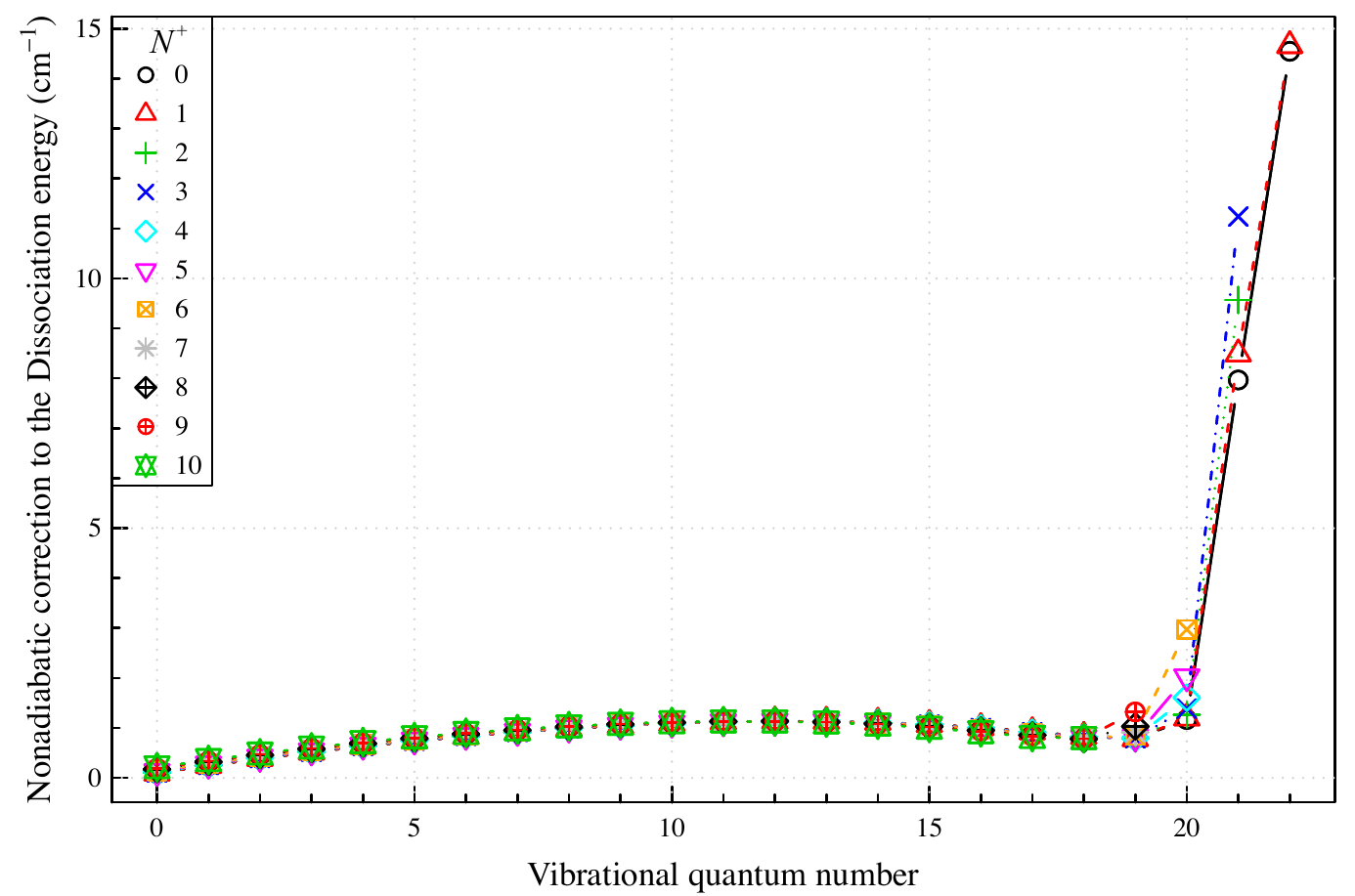}\\
	\caption{Nonadiabatic corrections to the dissociation energies of the bound states of HD$^+$ obtained as differences between the adiabatic energies calculated in the present work and the nonadiabatic energies calculated by Moss \cite{moss93b}. The lines connecting the data points are used to guide the eye.
		\label{CRHD:fig:calc_eng}}
\end{figure}

To assess the quality of the calculated Franck-Condon factors and to explore the effects of the {\it g/u} mixing on the level positions of the bound states of HD$^+$, we calculated them based on Eq.~(\ref{H2+_hamiltonian}) in the adiabatic approximation using the methods described in Refs.~\cite{beyer16a,beyer17a}. The comparison with the nonadiabatic dissociation energies reported by Moss \cite{moss93b} is presented in Fig.~\ref{CRHD:fig:calc_eng}, which displays the differences between our values and those calculated by Moss. These differences correspond to the nonadiabatic corrections. Up to $v^+=18$ the nonadiabatic corrections are comparable to the corrections obtained in H$_2^+$ and D$_2^+$ (see Table 2 of Ref.~\cite{beyer16b} and Tables 1 and 2 of Ref.~\cite{beyer17a}). In this case, the adiabatic potential-energy function of the X$^+$ state can be used to obtain reliable Franck-Condon factors. Whereas the nonadiabatic corrections almost vanish for the highest vibrational states of H$_2^+$ and D$_2^+$, they increase very rapidly beyond $v^+=18$ in the case of HD$^+$. This behaviour is attributed to the fact that the single adiabatic potential function of the X$^+$ state fails to describe the distinct H$^+$ + D(1s) and H(1s) + D$^+$ dissociation thresholds and has the same origin ({\it g/u} mixing) as the gradual evolution toward equal intensities of transitions to even- and odd-$N^+$ levels with increasing $v^+$ values discussed in the context of Figs.~\ref{CRHD:fig:spec_HH_zeke} and \ref{CRHD:fig:spec_HH_BB_zeke}.

%
%
\subsection{Resonances}

The PFI-ZEKE PE spectra recorded from the $\bar{\rm H}$(11,4) and $\bar{\rm B}$(11,2) intermediate states and depicted in Fig.~\ref{CRHD:fig:spec_HH_BB_zeke} also have structure in the dissociation continuum. In both cases, an increase of the signal is observed just above the first dissociation threshold (H$^+$ + D(1s)). Both spectra also show a sharp feature just above each of the two (H$^+$ + D(1s) and D$^+$ + H(1s)) thresholds, which we assign to the X$^+(21,4)$ and A$^+(0,4)$ quasibound states, respectively. The sharp feature marked with an asterisk in the upper panel of Fig.~\ref{CRHD:fig:spec_HH_BB_zeke} was found \emph{not} to be a feature of the continuum but can be unambiguously identified as one of the spurious transitions discussed in Subsection~\ref{sec:bound}. 

The spectrum displayed in the upper panel of Fig.~\ref{CRHD:fig:spec_HH_BB_zeke} reveals a weak, broad feature that we attribute to the X$^+(20,7)$ quasibound state. This feature is not visible in the lower panel because of the gerade symmetry of the H$\bar{\rm H}$ state and the lower intensity of transitions to rotational states odd $N^+$ values of when starting from even-$N$ rotational levels of the $\bar{\rm H}$ state.

\begin{figure}\centering
	\includegraphics[width=0.8\textheight,angle=90]{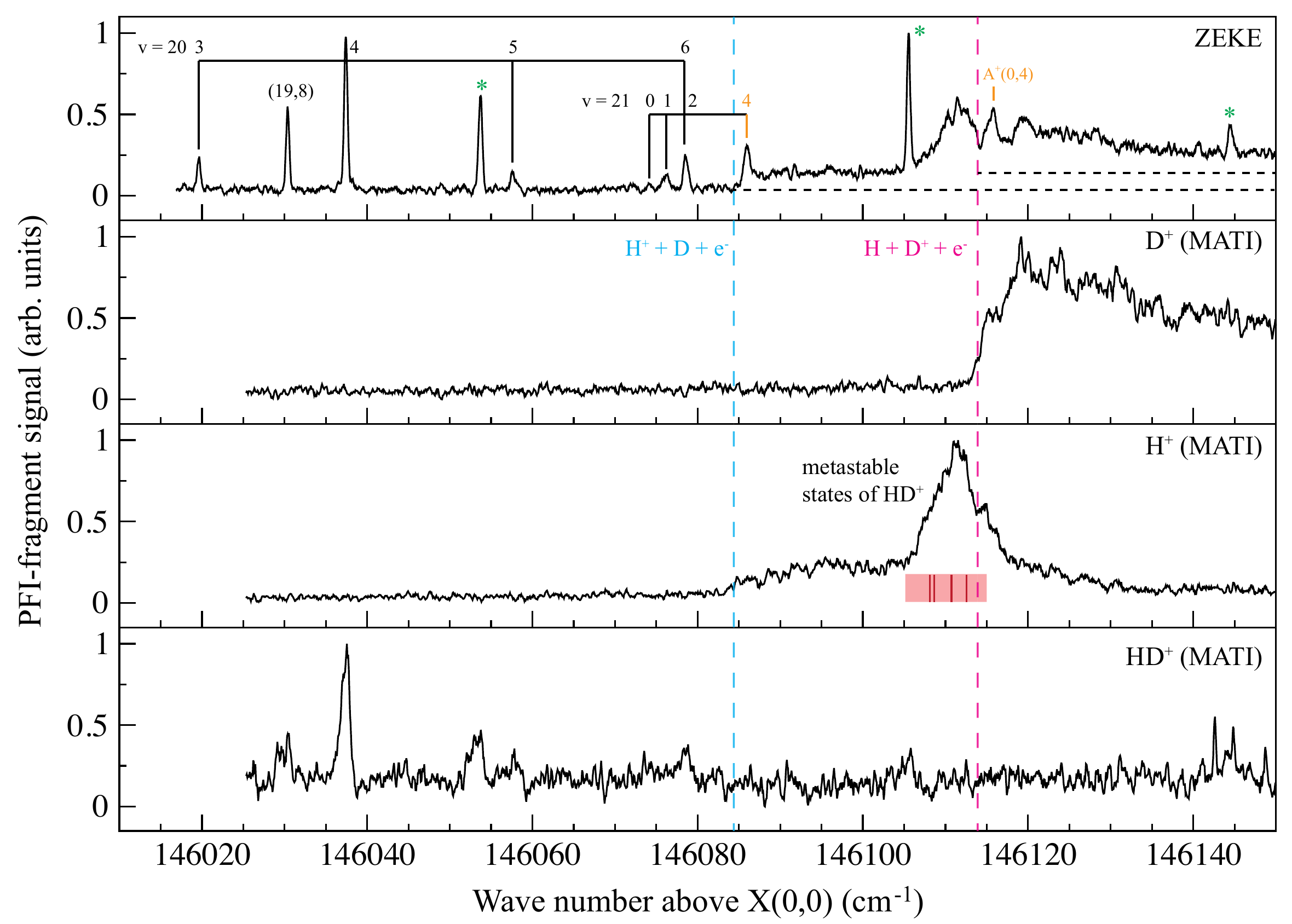}
	\caption{PFI-ZEKE photoelectron (top panel) and MATI (2nd panel: D$^+$; 3th panel: H$^+$; bottom panel: HD$^+$) spectra of HD recorded via the $\bar{\rm H}$(11,2) intermediate state. The first and second dissociation thresholds are indicated by blue and red dashed lines, respectively. The calculated level positions of the A$^+$ metastable states from Ref.~\cite{wolniewicz91a} are indicated by red bars and their calculated overlapping widths by a red rectangle.} \label{CRHD:fig:spec_res}
\end{figure}

The spectrum in the lower panel of Fig.~\ref{CRHD:fig:spec_HH_BB_zeke} reveals a second broad resonance just below the second dissociation threshold (D$^+$ + H(1s)) with a sharp decrease in signal intensity at the dissociation threshold.
To elucidate the origin of this resonance, we recorded MATI spectra by monitoring the HD$^+$, D$^+$ and H$^+$ signals resulting from the delayed pulsed-field ionisation in separate time-of-flight windows. These spectra are compared with the corresponding PFI-ZEKE PE spectrum in Fig.~\ref{CRHD:fig:spec_res}. Below the first dissociation threshold (H$^+$ + D(1s)), the lines observed in the PFI-ZEKE spectrum all appear in the HD$^+$ MATI spectrum and stem from the pulsed-field ionisation of high-$n$ molecular Rydberg states of HD.
Above the first threshold, the signal in the H$^+$ channel increases and corresponds to the field ionisation of H($nl$) Rydberg states produced by dissociation. Above the H$^+$ + D(1s) dissociation threshold, the HD$^+$ ion core thus dissociates and the Rydberg electron follows the charged fragment (here H$^+$). The Rydberg electron thus acts as a spectator to the HD$^+$-core dissociation. The broad feature that is observed in the PFI-ZEKE PE spectrum just below the second threshold also appears solely in the H$^+$ channel. Based on the calculations of Wolniewicz and Orlikowski \cite{wolniewicz91a}, we attribute this broad line to the overlapping A$^+(0,0$-$3)$ Feshbach resonances marked by red bars below the H$^+$ MATI spectrum . A nonzero signal is only observed in the D$^+$ channel above the second dissociation threshold (H(1s) + D$^+$) corresponding to excitation of molecular Rydberg states that dissociate into H(1s) and D($nl$).

\begin{table} \centering
	\caption{\small Observed positions and widths of the quasibound states in HD$^+$ above to the H$^+$ + D(1s) dissociation threshold. The relative experimental values were converted to absolute energies using the dissociation energy of the X$^+(20,4)$ level reported by Moss \cite{moss93b} (46.9909~cm$^{-1}$) (all values in cm$^{-1}$). \label{CRHD:tab:res_engs}}
	\begin{tabular}{lrrrr}
		\toprule
		Level        & \multicolumn{2}{c}{Observed} &                  \multicolumn{2}{c}{Calculated}                     \\[2pt]\cline{2-3}\cline{4-5}\\[-2pt]
		&  Position &            Width &                           Position &                      Width   \\ \midrule
		X$^+$(20,7)  &  11.18(9) &           2.1(3) &                 12 \cite{davis78a} &                         --   \\
		X$^+$(21,4)  & 1.656(19) &          0.28(7) &               1.55 \cite{davis78a} &                         --   \\
		A$^+$(0,0-3) &     23(1)-29(1) &               -- & 23.975-28.365 \cite{wolniewicz91a} &                   --           \\
		A$^+$(0,4)   &  31.19(3) &           0.8(2) &        31.076 \cite{orlikowski94a} & 0.567 \cite{orlikowski94a}   \\ \bottomrule
	\end{tabular}
\end{table}

The resonance widths and positions observed experimentally are given in Table~\ref{CRHD:tab:res_engs}. They agree less well with the calculations than was the case in H$_2^+$ \cite{beyer16a,beyer16b} and D$_2^+$ \cite{beyer17a}. The agreement is nevertheless sufficient for these assignments to be made with some confidence. The assignments of (i) the A$^+(0,4)$ resonance as shape resonance with H(1s) + D$^+$ character and (ii) the A$^+(0,0$-$3)$ resonances as Feshbach resonances of H(1s) + D$^+$ character, and (iii) the X$^+(21,4)$ and X$^+(20,7)$ resonances as shape resonances of H$^+$ + D(1s) character lend support to the classification of these resonances proposed by Davis and Thorson \cite{davis78a}. 

\section{Conclusions}

In this article, we have presented a study of the structure and nonadiabatic dynamics in HD and HD$^+$ resulting from the breakdown of the {\it g/u} symmetry at long range. The observations concerned the level structure and dissociation dynamics of HD in the $\bar{\rm H}$ and $\bar{\rm B}$ outer-well states into H$(n=2)$ + D(1s) and H(1s) + D$(n=2)$ fragments and of HD$^+$ near the dissociative-ionisation threhsolds H$^+$ + D(1s) and H(1s) + D$^+$. 

The preferential formation of H$(n=2)$ and D$(n=2)$ fragments in the dissociation of the $\bar{\rm B}$ and $\bar{\rm H}$ states, respectively, was interpreted as originating from the dominant role of one-electron-transfer processes over two electron processes [see Eqs.~(\ref{CRHD:eq:pred_B}) and~(\ref{CRHD:eq:pred_H})]

The determination of precise X$^+$ rovibrational level energies for the weakly bound states of HD$^+$ just below the H$^+$ + D(1s) and the comparison with adiabatic and nonadiabatic levels energies enabled the observation of a sharp onset of nonadiabatic corrections beyond $v^+=18$ caused by {\it g/u}-symmetry breaking.
The separate identification of the H$^+$ + D(1s) and H(1s) + D$^+$ dissociation thresholds in the middle panels of Fig.~\ref{CRHD:fig:spec_res} enabled the determination of the relative positions of these thresholds (see also Ref.~\cite{beyer18c}). The mass selectivity of the MATI spectra also permitted the unambiguous attribution of the spectral features observed in the continuum to either of the two dissociative ionisation channels.

The comparison of the dissociative photoionisation cross sections observed in the spectra recorded from the H$\bar{\rm H}$ and B$^{\prime\prime}\bar{\rm B}$ state in Fig.~\ref{CRHD:fig:spec_HH_BB_zeke} leads to the following conclusions:
\begin{itemize}
	\item The spectrum recorded from the $\bar{\rm B}$(11,2) state reveals a large step at the H$^+$ + D(1s) threshold but hardly any further increase of signal at the D$^+$ + H(1s) dissociation threshold. The D$^+$ + H(1s) channel is thus dark in this excitation sequence.
	\item In contrast, the spectra recorded from the $\bar{\rm H}$(11,4) state show a very weak step at the H$^+$ + D(1s) threshold and a large increase at the D$^+$ + H(1s) threshold. It is now the H$^+$ + D(1s) channel that is dark in this excitation sequence.
	\item The A$^+$ Feshbach resonances are not observed in spectra recorded from the $\bar{\rm B}$(11,2) level, because they are associated with the D$^+$ + H(1s) channel, which is dark. The Feshbach resonances are detected in the H$^+$ signal because of electronic predissociation into the H$^+$ + D(1s) continuum. 
\end{itemize}

This behaviour is particularly interesting: Both the intermediate $\bar{\rm H}$ and $\bar{\rm B}$  states and the final ionic X$^+$ and A$^+$ states are of mixed {\it g/u} character, resulting in a partial (in the case of the intermediate states) or complete (in the case of the final states in the dissociation continua) localisation of the electron(s) on either the proton or the deuteron. 

The dark dissociative-ionisation channels are the channels for which the two electrons would need to be excited, and the bright dissociative-ionisation channels those for which charge-transfer excitation takes place, which is a one-electron process [see Eqs.~(\ref{CRHD:eq:pred_B}) and ~(\ref{CRHD:eq:pred_H})].
These considerations are thus analogous to those used to explain the dissociation dynamics of the $\bar{\rm H}$ and $\bar{\rm B}$. 

\begin{figure}\centering
	\includegraphics[width=0.7\textwidth]{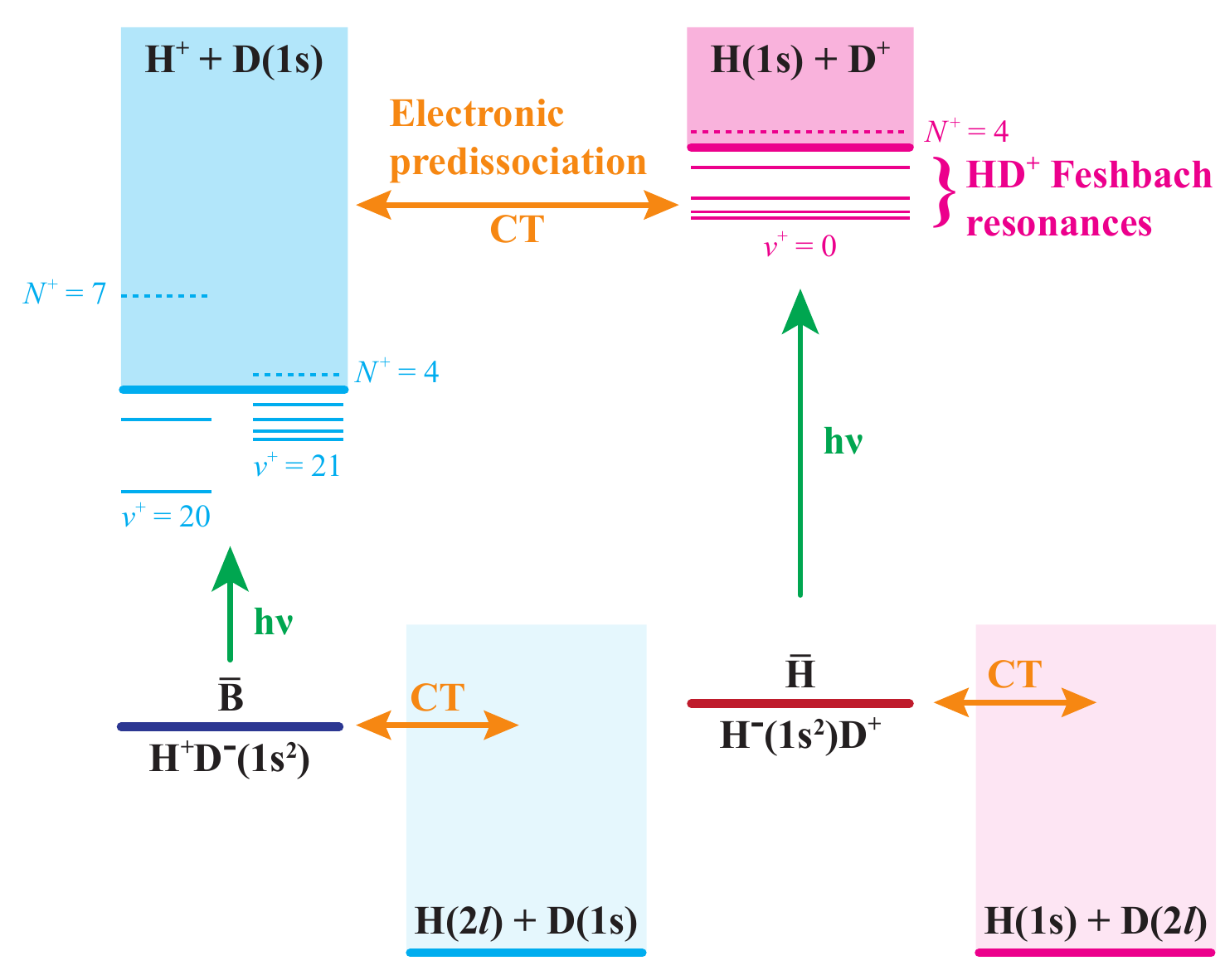}
	\caption{Schematic illustration of the dissociation dynamics in $\bar{\rm H}$ and $\bar{\rm B}$ states of HD and in the X$^+$ and A$^+$ states of HD$^+$. The upper blue and magenta level schemes correspond to the weakly bound rovibrational levels of HD$^+$ and the dissociative-ionisation continua associated with the H$^+$ + D(1s) and H(1s) + D$^+$ dissociation thresholds, respectively,which correlate adiabatically to the X$^+$ and A$^+$ electronic states of HD$^+$. Resonances that can be classified as shape and Feshbach resonances are drawn as dashed and full horizontal lines, respectively. The lower pale blue and magenta dissociation continua are associated with the H(2$l$) + D(1s) and H(1s) and D(2$l$) dissociation continua of HD. The green and orange arrows indicate the bright one-electron charge-transfer excitation and the one-electron charge-transfer predissociation, respectively.  See text for details. \label{CRHD:fig:HD_dynamics_scheme}}
\end{figure}

The conclusions drawn in the present study concerning the dissociation dynamics of the $\bar{\rm H}$ and $\bar{\rm B}$ states in HD, of the X$^+$ and A$^+$ levels in HD$^+$, and the intensities in the photoionisation and photoelectron spectra are all related to nonadiabatic {\it g/u}-symmetry breaking and are summarised schematically in Fig.~\ref{CRHD:fig:HD_dynamics_scheme}.

The results presented in this article illustrate the fact that {\it g/u}-symmetry breaking makes the dissociative-ionisation dynamics in HD is richer than in H$_2$ and D$_2$. Accurate calculations of the resonances in the dissociation continua of HD$^+$ observed for the first time in the present work would be desirable to confirm our assignments.

\section*{Acknowledgements}
This work is supported financially by the Swiss National Science Foundation (Grant No. 200020B-200478) and the European Research Council through an advanced grant under the European Union's Horizon 2020 research and innovation programme (Grant No. 743121).

\bibliographystyle{tfo}

\end{document}